\documentclass[journal,12pt,onecolumn,draftclsnofoot,]{IEEEtran}
\IEEEoverridecommandlockouts
\usepackage{cite}
\usepackage{verbatim}
\usepackage{listings} 
\usepackage{listings, xcolor}

\definecolor{verylightgray}{rgb}{.97,.97,.97}

\lstdefinelanguage{Solidity}{
	keywords=[1]{anonymous, assembly, assert, balance, break, call, callcode, case, catch, class, constant, continue, constructor, contract, debugger, default, delegatecall, delete, do, else, emit, event, experimental, export, external, false, finally, for, function, gas, if, implements, import, in, indexed, instanceof, interface, internal, is, length, library, log0, log1, log2, log3, log4, memory, modifier, new, payable, pragma, private, protected, public, pure, push, require, return, returns, revert, selfdestruct, send, solidity, storage, struct, suicide, super, switch, then, this, throw, transfer, true, try, typeof, using, value, view, while, with, addmod, ecrecover, keccak256, mulmod, ripemd160, sha256, sha3}, % generic keywords including crypto operations
	keywordstyle=[1]\color{blue}\bfseries,
	keywords=[2]{address, bool, byte, bytes, bytes1, bytes2, bytes3, bytes4, bytes5, bytes6, bytes7, bytes8, bytes9, bytes10, bytes11, bytes12, bytes13, bytes14, bytes15, bytes16, bytes17, bytes18, bytes19, bytes20, bytes21, bytes22, bytes23, bytes24, bytes25, bytes26, bytes27, bytes28, bytes29, bytes30, bytes31, bytes32, enum, int, int8, int16, int24, int32, int40, int48, int56, int64, int72, int80, int88, int96, int104, int112, int120, int128, int136, int144, int152, int160, int168, int176, int184, int192, int200, int208, int216, int224, int232, int240, int248, int256, mapping, string, uint, uint8, uint16, uint24, uint32, uint40, uint48, uint56, uint64, uint72, uint80, uint88, uint96, uint104, uint112, uint120, uint128, uint136, uint144, uint152, uint160, uint168, uint176, uint184, uint192, uint200, uint208, uint216, uint224, uint232, uint240, uint248, uint256, var, void, ether, finney, szabo, wei, days, hours, minutes, seconds, weeks, years},	% types; money and time units
	keywordstyle=[2]\color{teal}\bfseries,
	keywords=[3]{block, blockhash, coinbase, difficulty, gaslimit, number, timestamp, msg, data, gas, sender, sig, value, now, tx, gasprice, origin},	% environment variables
	keywordstyle=[3]\color{violet}\bfseries,
	identifierstyle=\color{black},
	sensitive=false,
	comment=[l]{//},
	morecomment=[s]{/*}{*/},
	commentstyle=\color{gray}\ttfamily,
	stringstyle=\color{red}\ttfamily,
	morestring=[b]',
	morestring=[b]"
}

\usepackage{amsmath,amssymb,amsfonts}
\usepackage{algorithmic}
\usepackage{graphicx}
\usepackage{textcomp}
\usepackage{xcolor}
\usepackage{hyperref}
\def\BibTeX{{\rm B\kern-.05em{\sc i\kern-.025em b}\kern-.08em
    T\kern-.1667em\lower.7ex\hbox{E}\kern-.125emX}}
\begin{document}

\iffalse
\title{IoT Survey *\\
{\footnotesize \textsuperscript{*}Note: Sub-titles are not captured in Xplore and
should not be used}
\thanks{Identify applicable funding agency here. If none, delete this.}
}
\fi
%\title{An application-based survey on IoT-Online Shopping with Steps, Integration and, Vulnerabilities of IoT Data and Mitigation with Blockchain}

\title{A Survey on the Applications of Blockchains in Security of IoT Systems}

%The Applications of Blockchains in Addressing the Integration and Security of IoT Systems: A Survey}

\author{\IEEEauthorblockN{Zulfiqar Ali Khan and Akbar Siami Namin}\\
\IEEEauthorblockA{\textit{Department of Computer Science} \\
\textit{Texas Tech University}\\
\textit{zulfi.khan@ttu.edu; akbar.namin@ttu.edu}}
}

\maketitle

\begin{abstract}
The Internet of Things (IoT) has already changed our daily lives by integrating smart devices together towards delivering high quality services to its clients. These devices when integrated together form a network through which massive amount of data can be produced, transferred, and shared. A critical concern is the security and integrity of such a complex platform to ensure the sustainability and reliability of these IoT-based systems. Blockchain is an emerging technology that has demonstrated its unique features and capabilities for different problems and application domains including IoT-based systems. This survey paper reviews the adaptation of Blockchain in the context of IoT to represent how this technology is capable of addressing the integration and security problems of devices connected to IoT systems.  The innovation of this survey is that we present a survey based upon the integration approaches and security issues of IoT data and discuss the role of Blockchain in connection with these issues.

%(TBC) The interaction of humans with IoT devices is increasing through wearable, health care, and online shopping devices. In the same way, Blockchain is also making progress in different spheres of human life independently or in conjunction with IoT devices. Hence forth
%users and novice researchers must become aware of the underlying principles, recent advancements, and Vulnerabilities of both the technologies. One approach for this familiarization is to provide a comprehensive but compact survey. 
%The innovation of our work is that we present a survey based upon the integration approaches and security issues of IoT data and discussed the role of Blockchain in connection with both these issues. We provided examples from the state-of-the art researches to highlight Blockchain-based uses cases. %We have identified the tools corresponding to each of the vulnerable categories.   
\end{abstract}

\begin{IEEEkeywords}
Internet of Things, Vulnerabilities, Blockchain, Classification of IoT Integration
\end{IEEEkeywords}

%\section{Introduction}

%TBD.

%TBD. I am having hard time understanding this paper. What integration means here?  

\vspace*{-0.08in}
\section{Introduction}
\label{Intro}

The devices connected to the Internet of Things (IoT) are capable of generating big amount of data captured by its sensors. Managing and handling this big data plays a vital role in quality and assurance of IoTs. IoT is the building block of smart objects \cite{IMF3} where these objects are communicating through a wireless network. Examples include  Radio  Frequency Identification (RFID)-based low battery-powered disposable computer chips having computational power. 

%which can cause storage and processing issues. IoT is the binding block of smart objects \cite{IMF3},\cite{IoTPR4} where smart objects are communicating through a wireless network. Examples include  Radio  Frequency Identification (RFID)-based-low battery-powered disposable computer chips or ``Things'' having computational power. 

%Therefore, managing and handling this big data plays a vital role in IoT. 
The concept of big data management can be observed in various contexts such as finance, medical care, and education, to name a few. For instance, in the application domain of financial or commercial application domains, it is essential to transfer financial data and transactions in a very secure way. This problem has created FinTech (Financial Technology) \cite{IMF25} where the objective is to use the latest innovation to make monetary transfers and financial services like online shopping (OLS) secure and affordable. 

Apart from FinTech's invention, financial sectors act as the backbone of the economy. Financial institutions are aware of their critical roles and subsequently respond to the market sentiments rapidly.  For instance, the banking sector started as bricks and mortar locations but gradually replaced several branches with Virtual Banking (i.e., e-Banking). However, when the financial gurus realized the importance of cash, they again focused on creating physical bank branches. On the other hand, physical units face issues of insecurity, unnecessary charges, and long queues. In most cases, the problems are so grave that they result in the capital flight. The solution lies in the adoption of IoT in the finance and online shopping domain. IoT provides tremendous opportunities for the growth of some critical application domains such as the financial sector by enhancing all three phases of the financial data flow system, namely: data retrieval, data analysis and management, and data reporting \cite{IMF27}. As a result, both traders and consumers have access to more information about goods instantly, monitor different asset markets, and make them available for better opportunities.

%Previous surveys provide state-of-the-art information on several related or unrelated topics. However, this approach does not help the developers who want to focus on an application development and are interested in exploring different steps of the application along with the current research on it. To solve this problem, we innovated the idea of application-based surveys and provided its details in this paper.
%previous IoT integration surveys??
%TBD.

Due to its key role in addressing critical issues of IoT-based systems, this paper surveys the applications of Blockchain in integration and security of IoTs. %The survey first presents the general architecture of typical IoTs. It then 
The paper first reviews research studies in integration of IoT data. We then present the remedial features that Blockchain offers in mitigating IoT's vulnerabilities. The key contributions of this paper are as follows:
\begin{itemize}
    \item[--] It introduces a classification of integration approaches in IoT systems.
    \item[--] It presents a survey of the applications of Blockchain technology to the problems related to the integration and security in IoT systems. 
    \item[--] It highlights the research gap and future application domains of the Blockchain technology in IoT contexts.
%    \item[--] We are the first to incorporate both the Cloud and Blockchain based integration techniques in our survey. 
\end{itemize}

The rest of this paper is organized as follows: Section \ref{sec:methodology} discusses the methodology taken in conducting this survey. Section \ref{sec:Integration}  presents our classification of integration in IoT systems. In Section 
\ref{Sec:IoTVulnerabilities}, we provide a survey on the applications of Blockchain in IoTs. Section \ref{sec:conclusion} concludes the paper and highlights the future research directions.

\vspace*{-0.08in}
\section{Methodology}
\label{sec:methodology}

The purpose of this survey paper is not to conduct any meta-analysis but review the most intriguing applications of Blockchain for addressing the problems unique to IoT systems. In preparation of this survey paper, we followed the general guidelines recommended by the Preferred Reporting Items for Systematic Review and Meta-Analyses (PRISMA) \cite{1011}. First, we started our data collection using the search term ``IoT Blockchain vulnerabilities''. Then, we identified and collected only the relevant articles retrieved by the Google search engine that were reported. We then archived and classified the collected articles and proceedings papers. We repeated the procedure on the university's portal and online library system searching for digital libraries of reputable publishers such as IEEE, ACM, science direct, Springer, MDPI, and Elsevier with the optional parameter indicating the year of publications.% (i.e., 2021).

\vspace*{-0.05in}
\section{Integration of IoT Data}
\label{sec:Integration}

IoT devices (e.g., sensors, IoT-enabled cameras) usually generate massive data. On the other hand, these IoT devices often lack resources to process or store the large amount of data captured and transmitted by sensors. This drawback of IoT devices justifies the needs for integration of IoT data. In other words, integration (for instance, with cloud computing and Blockchain) provides the required memory and processing power for storing and organizing the IoT data for machine learning and AI applications to make decisions. However, keeping data in cloud or Blockchain requires transmission of data from an IoT network to cloud (or Blockchain network), which may cause the stored data to incorporate noise or lose its privacy due to interception by the attacker. The data encryption mechanism can handle the privacy issue. 

Saxena et al.\ \cite{imp11} express the concerns of data corruption related to the sensor data stored in the Blockchain. This problem requires that the sensors retrieving the stored data from the Blockchain (or Cloud) must verify the data that another sensor might have generated. Wang et al.\ \cite{ Slot21} introduce an approach for verifying the sensor data by implementing a collaborative Certificate Validation protocol. \iffalse and a mechanism for verifying the validated certificate corresponding to the data\fi The technique uses the cache space on all IoT devices to create a large memory pool for storing validated certificates. IoT device, which retrieves the data from the cloud, needs to verify the authenticity of data before usage. The device retrieves the cached certificate corresponding to the retrieved data and asks the certificate generator (i.e., holder) to verify it.  %After verification, if the data changes, then the protocol revokes its certificate. An AI engine then uses the stored data to make decisions. 

Based on our survey findings, several integration techniques are incorporating IoT data for processing and/or  storage such as 1) standalone (i.e., sensors have no interaction with other systems), 2) integrated (i.e., sensors interact either with Cloud or Blockchain systems) 3) collaborative (i.e., sensors interact with one or more technologies like cloud, Blockchain, or Fog) and 4) Adhoc approaches (i.e., temporary availability of facilities like cloud computing for integration of IoT data.) \iffalse where\\ collaborative: means more than one system like Cloud and (or) Blockchain associated with storage and processing of IoT data.\fi 
\iffalse adhoc: means temporary availability of facilities for integration of IoT data.\fi

\iffalse
Figure \ref{fig:IoTDataIntegration} shows our classification of these approaches. In this survey paper, we provide a thorough discussion on the Blockchain integration and a few examples of other approaches from the latest research.

\begin{figure}[h]
  \centering
  \includegraphics[width=0.8\linewidth]{Classification of IoT Data Integration.png}
  \caption{Integration Techniques:Standalone, Blockchain and Cloud based, Colloborative and Other forms.}
  \label{fig:IoTDataIntegration}
\end{figure}
\fi

%\vspace*{-0.08in}
\subsection{Standalone Integration (No Integration)}\label{Inte:Stan}

%TBD. (I need to know the reference below so I can write better description).
%\cite{Slot21}
%Research Question\#2:\textcolor{orange}{How sensors process their data without integrating itself with Cloud or Blockchain?} 

This is a special case where each IoT device works independently. As a result, the data captured by its sensor are not transmitted or shared with some other devices. Consider a scenario in which IoT network does not connect to the cloud, Blockchain, or any other system for processing and streamlining the data by itself.

\iffalse
as shown in the Figure \ref{fig:CombinedCache} for the storage of validated certificates.

\begin{figure}[h]
  \centering
  \includegraphics[width=0.8\linewidth]{combining cache of all sensors.png}
  \caption{Combining cache memory: Sensor\(_1\) to Sensor\(_N\).}
  \label{fig:CombinedCache}
\end{figure}

\fi

Wang et al.\ \cite{Slot21} provide an example where the sensors use the cache space on all IoT devices to create a large memory pool for the storage of validated certifications. When an IoT device receives public key certificates from the cloud storage, it can update or revoke the certificate in the shared data. For this purpose, the IoT device has to locate the device (i.e., holder) that cached and verified the certificate. %Their work designed a fast locator to find the holders using Othello hashing\cite{1013} which consumes less memory as compared to the lookup tables and applies to programmable networks.

\subsection{Integrated Approaches}\label{Integ:InteAppr}

%TBD. A similar situation here, I need to know the reference given below so I can provide a better description.
We divide these approaches into two types: Cloud-based integration and Blockchain-based integration.

\subsubsection{Cloud-based Integration}\label{Inte:CloudbasedInte}
Zhao et al.\ \cite{Slot10} created a wireless monitoring system using a commercial device that sends data to the cloud using a low-power, long-range LoRaWAN gateway. The authors installed the unit on the existing dispenser on their campus and used a holder to monitor dispenser interaction. Supplied data helps the cloud service to support the unit staff with decision-making like refilling the dispenser.

\subsubsection{Blockchain-based Integration in IoT Systems}
\label{InteofIoTD:BCbasedInte}
Blockchain is an emerging technology that enables unlimited and decentralized data structure and processing. Due to its unique features, Blockchain can be utilized as a platform for addressing the concerns about vulnerabilities in IoT systems.  
%Blockchain became famous because of its unlimited and decentralized data structure. Blockchain provides a novel and low-cost mechanism for FinTech, but in IoT, Blockchain acts as a remedial for Vulnerabilities and a mechanism for storage cum processing of IoT data. First we would briefly discuss types of Blockchains.
We classify the IoT-Blockchain merger endeavors into Bitcoin, Ether, and HyperLedger types. It is also possible to extend the classification to include some other Blockchains technology. %Previous work's \cite{IMP11} integration technique focused on IoT, but this section introduces Blockchain-based integration techniques.

\paragraph{Bitcoin-based Integration}
\label{InteofIoTD:BCbasedInte:ClassofBCbasedInte:BitCTI}
This type of integration employs Bitcoin-based Blockchain for storing IoT data. For example, Ren et al.\ \cite{Slot9} use a new digital signature called sequential aggregate signature scheme with a designated verifier (DVSSA) on the IoT data received from the cloud. The system then stores the signed blocks into the Blockchain. The advantage of using this technology is data integrity, non-repudiation, and reduced size of messages. Some other research work such as Ren et al.\ \cite{Slot8} reward the miners with incentives, save power, and replace Proof of Work (PoW) consensus algorithm with provable data possession. The consensus algorithm ensures the agreement of a majority of miners to add a block to the Blockchain.

\paragraph{Ethereum-based Integration}
\label{InteofIoTD:BCbasedInte:ClassofBCbasedInte:EtheTI}

This type of integration involves the use of Smart Contracts (SCs).  SCs require crypto asset-based accounts to store digital assets in the form of crypto-currencies like Ethereum \cite{IoTBC2}. This may be considered as a drawback because of the apprehension of attacks by hackers to steal Ether from the said accounts. One such attack is the DAO attack which exploits the reentrancy vulnerability as discussed in \cite{1010}.  

Blockchain provides a distributed and decentralized eco-system to maintain the security and privacy of stored \iffalse IoT-OLS\fi transactions (e.g., IoT-based online shopping systems (IoT-OLS)). Guo et al.\  \cite{IoTBC5} present an example of trading companies using smart meters and SCs to automate the payment process. The smart meter verifies the receipt of power and forwards the meter reading to SC. Once the SC is executed, it deducts the amount from the SC account and stores the payment into the power generating company’s account. 

\paragraph{HyperLedger-based Integration}
\label{InteofIoTD:BCbasedInte:ClassofBCbasedInte:HLTI}

Traceability of items \cite{IoTBC6} is an essential aspect of IoT-OLS activity. Recently, advancements in IoT-Blockchain integration allowed the IoT-OLS users to know the origin and state of food items and their various stages, from growing to consuming them \cite{IoTBC22}. Furthermore, the traceability aspect incorporates information assigned to the product related to its identity, processing, and usefulness \cite{IoTBC8}. The by-product of this endeavor is the buildup of consumers' trust in IoT-OLS. Furthermore, the added security aspect has invited the attention of several governments to create Blockchain Food Traceability Systems (BFTS) \cite{IoTBC8}. SC solutions can track the food chain from the fields to their transportation to the OLS stores as finished products and finally to the households as grocery items.
Thus, the homeowners can easily trace back to the farmers growing the food items.  For example, Garaus and  Treiblmaier \cite{IoTBC6} discuss the use of mobile phones to track the flow of food items with the help of scannable codes; whereas, Balamurugan et al.\ \cite{IoTBC16} present a system that uses IoT-HyperLedger Blockchain driven traceability techniques to record the current state of the food items retrieved through cryptographic IoT devices. 

The traceability technique in \cite{IoTBC10} uses Blockchain QR codes but does not specify the role of ChainCode (i.e., SC).
The study tries to address the concerns of Chinese consumers towards the use of dairy products in the context of 1) traceable safe foods and 2) the willingness to use QR codes.
 On the other hand, the HyperLedger framework presented by Anithal \cite{IoTBC21} uses ChainCode for traceability and to perform transactions like the registration of food items by the farmer, moving the items from the farmer’s site to the processor’s site by the distributor, and so on\iffalse to name a few\fi.

The HyperLedger fabric incorporates ChainCode that sends/receives notifications to/from participants and monitors any violations. Thus ChainCode can help monitor IoT devices in managing configuration files stored on the Blockchain\cite{IoTHS1}. When the local configuration file changes due to an attack, IoT devices recover the configuration files from the Blockchain. The ChainCode receives a notification about the successful download and records this information on the Blockchain for the administrator.
\subsubsection{Collaborative Integration}

These approaches depend upon the collaboration of cloud computing and/or Blockchain with other technologies including fog, Edge, AI, and machine learning. As an example of cloud and fog computing-based collaboration, the fog and cloud computing offer a collaborative environment through which Complex Event Processing (CEP) can be accomplished in real time for IoT applications \cite{IoTCI7}. The architecture offered by fog computing enables holding CEP-based applications on low-cost resources. The architecture helps in generating automatic alarms when a relevant situation or condition occurs in the underlying IoT system. In addition, distributed offloading from cloud to fog architectures offers load balancing and thus reduces latency.

%The work [IoTCI7] focuses on fog and cloud computing collaboration in processing Complex Event Processing (CEP) based real-time IoT applications. IoT devices generate a large amount of heterogeneous data. Fog architecture holds CEP applications on low-cost resources, which generate alarms when a relevant situation occurs. In addition, distributed offloading from cloud to fog architectures provides load balancing and reduces latency.

\subsection{Adhoc Integration}
Uddin et al.\  \cite{IoTCI20} use ad-hoc approaches in agricultural lands presenting its application in far-off places with no internet access. Large IoT-OLS stores set up in similar locations can benefit from these approaches. The major drawback in such situations is that an immediate cloud or Blockchain integration may not be feasible. For instance, Uddin et al.\  \cite{IoTCI20} present a system that uses unmanned aerial vehicles (i.e., UAVs similar to drones). The vehicle brings the cloud to the proximity of agricultural lands thus implementing edge computing. UAV itself handles the field location problem without the aid of GPS. IoT devices communicate with UAV using short-range protocols like Zigbee and Bluetooth. On the other hand, WiMax facilitates WAN communication between the UAV and the remote cloud network. Thus, UAV solves the local requirements of the sensors, whereas the remote cloud solves the global requests of the UAV.

\section{IoT Vulnerabilities and Blockchain Remedial}
\label{Sec:IoTVulnerabilities}

Vulnerability in IoTs \cite{IoTVul5} is a security glitch in an executable system that can weaken a system's sovereignty so much that an attacker misuses the owner's or its organization's resources for pleasure or personal profits. In IoT, the vulnerability problem is serious because the technology empowers IoT devices to operate independently. The defense lies in the offline execution of critical systems. For instance, car manufacturers now equip vehicles with computers to handle collision avoidance.  However, if the a hacker gains access to these vital computers, the car's operations can be at stake.% \cite{IoTRisk1}.

Adversaries can also affect IoT deployments through unintended physical interactions caused by applications. These applications  control the actuators and sensors. They may work fine in isolation, but in an integrated environment, they may land the devices in an unsafe state due to unexpected interactions \cite{IoTVul6} (e.g., controlling the pressure cooker and vacuum cleaner).

Khan and Salah \cite{imp22} concentrate on nineteen IoT vulnerabilities. Here, we focus on the vulnerabilities discussed in \cite{IoTVul4, IoTVul1}. Incidentally, we found commonality between the discussed vulnerabilities in \cite{IoTVul4} and \cite{IoTVul1}. This survey concentrates on the vulnerabilities in  \cite{IoTVul1} but readers can refer to  \cite{IoTVul4} for additional details. The uniqueness of our work is that we provide Blockchain-based remedial for each of the discussed vulnerabilities. Blockchain has several key features that can better address some of the vulnerabilities in IoT. Khan and Namin \cite{1010} discuss some of the vulnerabilities of Blockchain in the context of SCs.

\subsection{Deficient Physical Security Controls}% (Lack of Physical Hardening}
\label{IoTVulBCR:DefPS}

Waheed et al.\ \cite{DPS5_mitifolder} discuss several physical threats related to IoT devices such as causing harm to devices by direct, environmental, power, or intrusion like interferences. Apart from this, there is no lifetime guarantee for the IoT devices’ components. The standard technique exploits weak passwords\cite{DPS3} to inject data tampering once the attacker gets access to the device. Otherwise, the attacker misuses the network traffic due to untrusted protocols and then steal specific sensor data. Furthermore, sensors are also vulnerable to firmware attacks occurring during the update process because IoT devices do not invoke authentication or encryption procedures \cite{1006}.

\subsubsection*{Blockchain Remedial}
\label{IoTVulBCR:DefPS:Remedial} 

Blockchain offers remedial to the device's physical security problems by maintaining a decentralized network protected by cryptographic keys. Each individual device can benefit from Blockchain’s general features by creating a Blockchain within its memory \cite{DPS2}. Thus the device keeps track of data updates by maintaining a ledger containing a history of previous transactions, preventing the risk of data tampering. Sensors exchanging their data for making valuable decisions require authentication along with the combination of %cum
key distribution techniques as discussed by Ding et al.\ \cite{IAA13}. SCs can also facilitate authenticated interaction, as discussed in \cite{DPS3} and the advantage is the revocation capability when the authentication period expires.

\subsection{Insufficient Energy Harvesting}
\label{IoTVulBCR:InsufEner}

Technical constraints render the IoT devices with small batteries insufficient to run computationally intensive security routines. An attacker exploits the situation by pouring the device with massive requests. Handling these requests deplete the device's energy, ultimately making the device inactive.

\subsubsection*{Blockchain Remedial}
\label{IoTVulBCR:InsufEner:Remedial} 

Existing research reports that Blockchain consumes very high energy, even more than the requirements of a developed country like Ireland \cite{IEH1}. However, recent research \cite{IoTBC14}, \cite{IEH7} shows that clustering can help reducing energy consumption for IoT-Blockchain integration. These research studies use a multi-layer structure that allows detecting compromised blocks through a consensus algorithm. Pajooh et al.\ \cite{IEH7} use a genetic algorithm for clustering incorporating attributes such as distance and energy. In these clustering, each cluster has a cluster head (CH) for data transmission where members do not communicate with each other. Therefore, for energy enhancement, the algorithm outputs fewer CH.

\subsection{Inadequate Authentications}% (Weak, guessable, or hardcoded passwords\cite{IoTVul4})}
\label{IoTVulBCR:InadAuth}

Authentication of IoT devices is crucial in IoT-OLS applications because embedded IoT devices \cite{IAA2} usually do not require user's credentials. In addition, due to resource limitations, IoT devices are less enthusiastic about leveraging complex encryption techniques to protect the data during transmission. Consequently, problems such as payment fraud are increasing. As a result, worried traders are promoting the idea of more than one form of authentication. %On breaking the authentication check, the attacker makes the financial data inaccessible to the user. 

\subsubsection*{Blockchain Remedial}
\label{subsubsec:IAuth} 

Remedial techniques focus on device identification and re-authentication. Malan et al.\ \cite{IAA8} report an authentication vulnerability incident when a user changed the password of his smart doorbell. Even though the new password was in place, the attacker was still able to access the doorbell using their mobile app. The reason was that the password change function did not ask for re-authentication of the device. In Blockchain each device has a public key for encrypting messages. 

Neshenko et al.\ \cite{IAA2} introduce the concept of developing a tempered proof universal identity of the device by combining the public key with International Mobile Equipment Identity (IMEI) and Original Equipment Manufacturer (OEM) firmware hashes. Gong et al.\ \cite{IAA12} combine the device's identity and weight and sends this information to a SC deployed in a Blockchain network of IoT gateways. All Blockchain copies validate the identity information. When a device connects to the Blockchain network, SC authenticates the device by verifying its ID.

\subsection{Improper Encryption}% (Insecure data transfer and storage\cite{IoTVul4})}
\label{IoTVulBCR:ImproperEnc}

Neshenko et al.\ \cite{IAA2} point out to the issue that IoT devices usually do not encrypt their local data, which can cause security problems. By local data we mean, configuration information, built-in settings, devices' passwords, etc. If the attacker gets physical access to the device, the attacker can alter the data mentioned above. But even if the user encrypts the local data using a weak encryption algorithm, one cannot rule out the chances of tampering \cite{IoTVBC4}.

\subsubsection*{Blockchain Remidal}
\label{IoTVulBCR:ImproperEnc:Remedial} 

Blockchain can provide a solution to the weak encryption (or no encryption) vulnerability of IoT devices by hashing the firmware and using the hash for the device authentication. Devices can have a universal identity by utilizing Blockchain’s public key. Therefore, a device can use the receiver’s identity for exchanging messages by encrypting the message and its hash. As a result, the receiver can decrypt the received package and validate the message by generating its hash 
and matching the new hash with the received hash. Otherwise, the sender and receiver can use Blockchain for communication. The latest endeavor performed by Loukil et al. \cite{IoTEn1}  discuss a privacy-preserving approach by employing homomorphic encryption and SCs for the consumers and an aggregator which encrypts the requested information. The  disadvantage is that the aggregation process coupled with encryption complicates the comprehension of information.

\subsection{Unnecessary Open Ports}% (Insecure network services, Lack of device management \cite{IoTVul4})}

The disadvantage of open ports is that they allow an attacker to access the device. Manufactures, by default, do not inform the consumer about the dangers of open ports \cite{IoTp1}. Open ports can have varied behaviors in the context of IPv4 and IPv6 \cite{IoTP4}. The open port vulnerability helped launching the Mirai attack which is   termed as the most dangerous IoT attack \cite{IoTPB6}. Mirai leveraged the telnet port and default credentials. Attackers might use a search engine like Shodan to identify the vulnerabilities (i.e., misconfigurations) on Internet-connected devices \cite{IoTP2}. Moreover, researchers warn about UPnP (Universal Plug and Play) because it provides remote access to the intruder to steal critical credentials \cite{IoTP5}. %For users, Wi-Fi Inspector \cite{IoTP3} checks weak credentials. 

Wazzan et al.\ \cite{IoTPB6} summarize the working of Botnet in three steps like scanning and infecting the vulnerable device until the device starts communication with the BotMaster (i.e., Attacker robot). At this point, the BotMaster installs malware on the device and tries to get hold of open ports; if not available, malware kills the associated process to lock its port. Finally, all the bots waiting in the pool launch a massive DDoS (distributed denial of Service) attack, which blocks the networking channels, making it impossible to connect. 

\subsubsection*{Blockchain Remedial} Falco et al.\ \cite{IoTPB1} developed a tool NeuroMesh to counter Botnets using a Botnet technology by incorporating Bitcoin protocol. The advantage of Bitcoin is that it is a safe protocol. This property provides a safe platform for NueroMesh to order its bots (i.e., NeuroNodes) commands. NueroMesh would scan the system at the installation time and declare all the running processes as safe (i.e., whitelisted). Now, if any process installs itself on the system, NeuroNode would kill it. NueroMesh spreads out its bots in different file systems, allowing the bots to escape being killed by the attacker installed on the system. At the same time, it monitors all the open ports.

\subsection{Insufficient Access Controls}% (Insecure Ecosystem Interfaces\cite{IoTVul4})}

%Access controls deal with the authorization of rights to access a resource. 
A system obtains information about resources to access using an access control list (ACL). Thus, if the attacker gets permission to alter the ACL, then this vulnerability can allow the adversary to get unrestricted access. However, if the data is publicly visible by default as in Ethereum Blockchain, this is an access violation but ensures transparency. On the other hand, the enormity of IoT devices makes the issue of access control more severe when there is an option of privilege commands that can disable or damage IoT devices \cite{IoTVAC11}.

\subsubsection*{Blockchain Remedial}
Algarni et al.\ \cite{IoTVAC6} use the concept of Blockchain Managers, governed by Mandatory Access Control policy header, to ensure secure authorization. The said policy allows Blockchain Managers to remove and add IoT devices and control transactions. In addition, Blockchain Managers contain agents like Miner, an internet-enabled device, accountable for handling communication within the Blockchain manager, and an authorization agent who informs the miner about the permissions of IoT devices. The authorization agent also communicates with the authentication agent to retrieve the identity of the connected device (or user) for security clearance and resource classification using Mandatory Access Control.

\vspace*{-0.05in}
\subsection {Improper Patch Management}% capabilities (Lack of secure update mechanism\cite{IoTVul4})}

IoT devices may operate through vulnerable firmware due to the lack of proper testing or update. This situation indicates that IoT devices need software patching. But at the same time, we must accept that IoT devices lack hardware capabilities. Ray et al.\ in \cite{ECP10} state that this duplicate problem puts IoT devices in an uncertain situation. Thus, authors in \cite{ECP10} suggest hardware patching, which would allow reconfiguring the hardware to deal with unexpected vulnerabilities with the help of FPGA (Field Programmable Gateway Array). This will help in configuring electronic circuits using a hardware description language) technology. 
\iffalse We suggest implementing hardware patching with the help of jumper settings. Yet, traditional solutions like antiviruses won’t work because the repeated memory scan would consume much energy.\fi The surveyed literature provides the following recommendations related to patching firmware of IoT devices:
\begin{itemize}
    \item[--] Patch installation is a complex procedure in the context of low-resource IoT devices. Thus device manufacturers should focus on small patches instead of large patch installations \cite{IoTPMC2}.
    \item[--] Client-server architecture is not a favorable mechanism for installing updates on billions of IoT devices due to a single point of failure \cite{IoTPMC4}. 
    \item[--] Blockchain mechanisms for distributing firmware updates are appealing because of avoiding the single point of failure and promotion of consensus \cite{IoTPMC4}.  
    \item[--] To increase people's trust, IoT device manufacturers can open-source firmware and its updates \cite{IoTPMC5}.

\end{itemize}

\subsubsection{Blockchain Remedial}
The latest work by Fukuda  and  Omote \cite{IoTPMC9} discuss a method for installation of firmware updates on IoT devices by focusing on vendors, distributors, and SCs. First, the vendor entity registers the IoT devices and provides the updated resource on the Blockchain. Then, to support decentralization, several distributor nodes on the Blockchain download the modified firmware and record their repository's IP address on the Blockchain. Later on, the IoT device downloads the firmware and notifies the distributor by its signature. Finally, the distributor informs the SC by a transaction about the device’s update activity to receive the reward. The advantage of the decentralized technique is that it provides incentives to the distributors, and IoT devices do not have to decrypt the files.

\iffalse
\section{Research Directions}
\label{sec:conclusion}
In this survey we observed the following directions for future research:
%TBD.
\begin{itemize}
%    \item[--] Falco et al.\ \cite{IoTPB1} provide a solution for handling botnet in the context of Bitcoin. Future research work can focus on addressing this problem in the context of SCs and cloud computing.
    
%    \item[--] There is already some research work in the direction of creating Blockchain within an IoT device \cite{DPS2}. However, replicating, i.e., providing a cloud mechanism within an individual IoT device, is still an open research problem.
    
    \item,[--] One resistance in adopting IoT solutions is that the battery power of sensors evaporates soon. For agricultural lands, clustering as discussed in \cite{IoTBC14} and \cite{IEH7} can help in power savings, but how we can reduce the energy demands of a small IoT network. 
    
    \item[--] We have found some research in the context of sensors for standalone processing of IoT data, but still, this area needs more analysis.
    
    \item[--] Work in \cite{ECP10} discusses  re-configurable IoT devices using FPGA technology. Future research can explore this option to handle the limitation of IoT devices.

\end{itemize}

\fi

\vspace*{-0.08in}
\section{Conclusion and Future Work}
\label{sec:conclusion}

Blockchain is an emerging technology for implementing decentralized applications. The technology also is useful in some other application domains such as security of IoT. One of the key features of Blockchain is the introduction of consensus protocols that make it possible to address tampering and security issues along with tracing the sources of actions. This survey paper reviews the applications of such a sophisticated technology in the context of IoT. We particularly listed possible vulnerabilities in IoT systems and provided the mitigation and solutions that the Blockchain technology can contribute to address these security problems. What makes this survey valuable is the diverse set of applications of Blockchain in IoTs. Certainly, the vulnerabilities and the mitigation offered by Blockchain are not limited to those discussed here and there is
a myriad number of other Blockchain's applications in this context that can be explored in future work.

\vspace*{-0.08in}
\section*{Acknowledgement}
This research work is supported by National Science Foundation (NSF) under Grant No. 1821560.

\bibliographystyle{IEEEtran}
\bibliography{sourcefile-bib}

\end{document}